# SMART HOME, SECURITY CONCERNS OF IOT

Alessandro Ecclesie Agazzi
*Department of Computing and Informatics*
*Bournemouth University*
Poole, United Kingdom
alessandro@ecclesieagazzi.com

***Abstract*— The IoT (Internet of Things) has become widely popular in the domestic environments. People are renewing their homes into smart homes; however, the privacy concerns of owning many Internet connected devices with always-on environmental sensors remain insufficiently addressed. Default and weak passwords, cheap materials and hardware, and unencrypted communication are identified as the principal threats and vulnerabilities of IoT devices.**

**Solutions and countermeasures are also provided: choosing a strong password, strong authentication mechanisms, check online databases of exposed or default credentials to mitigate the first threat; a selection of smart home devices from reputable companies and the implementation of the SDN for the DoS/DDoS threat; and finally IDS, HTTPS protocol and VPN for eavesdropping.**

**The paper concludes dealing with a further challenge, "the lack of technical support", by which an auto-configuration approach should be analysed; this could both ease the installation/maintenance and enhance the security in the self configuration step of Smart Home devices.**

## I. INTRODUCTION

The IoT (Internet of Things) has become widely popular in the domestic environments. People are renewing their homes into smart homes (Zeng, Mare and Roesner, 2017), adopting both "hi-tech" items (smart voice assistant, security cameras, fitness trackers) and classic household objects (smart fridges, doorbells and heating) (Ncsc.gov.uk, 2019).

Smart-home device has been defined as "any single-purpose Internet-connected device intended for the domestic usage (such as thermostat, lightening, or blood-pressure monitor) or a hub-like device that connects and controls multiple single purpose devices (e.g. a Samsung SmartThings hub or Amazon Alexa)" (Apthorpe, Reisman and Feamster, 2019).

Although they make our life easier and smarter, domotics and IoT raise significant privacy concerns. While, in the past, people's online activity was restricted to web browsing, now, smart home devices' always-on sensors transmit information regarding the users' offline activities on the Internet (Apthorpe et al., 2019).

As a result, "The contents, patterns, and metadata of network traffic, flowing from and to these IoT devices, can all reveal sensitive information about a user's online activity." (Apthorpe, Reisman and Feamster, 2019). This extensive information may be valuable in several and different contexts, such as advertising and business intelligence (Apthorpe et al., 2019).

This paper will review the impact of the IoT devices in homes and it will focus on the following objectives (Enisa.europa.eu, 2020):

- Analysis of the implementation of IoT devices throughout the house and their cybersecurity challenges.
- Mapping threats and vulnerabilities in relations to IoT assets and considerations of countermeasures.
- Giving an overview of techniques which may endanger IoT devices.

## II. SECURITY VULNERABILITIES AND THREATS

Smart home devices cannot be considered as specialised machines with built-in intelligence but as real computers that perform specialised tasks; for instance, a smart refrigerator can be seen as a computer which keeps track of its content and temperature or a smart lamp as a computer which can be switched on and off, remotely, by a smartphone. These tasks-specialised devices, known as IoT, smart home devices or domotics, are, almost always, composed by a microprocessor, like a laptop, connected to the internet (Angrishi, 2017).

Unlike specialised computers, however, IoT devices are very often developed and designed with a poor or even none security (Angrishi, 2017).

In addition, as Rapid7 explained (Stanislav and Beardsley, 2015), "IoT devices, unlike traditional computers, usually lack a proper update and upgrade patch once they leave the manufacturers' warehouse".

In the next section, vulnerabilities and threats will be analysed for a generic smart home device.

Here below, the data flow for a generic IoT device:

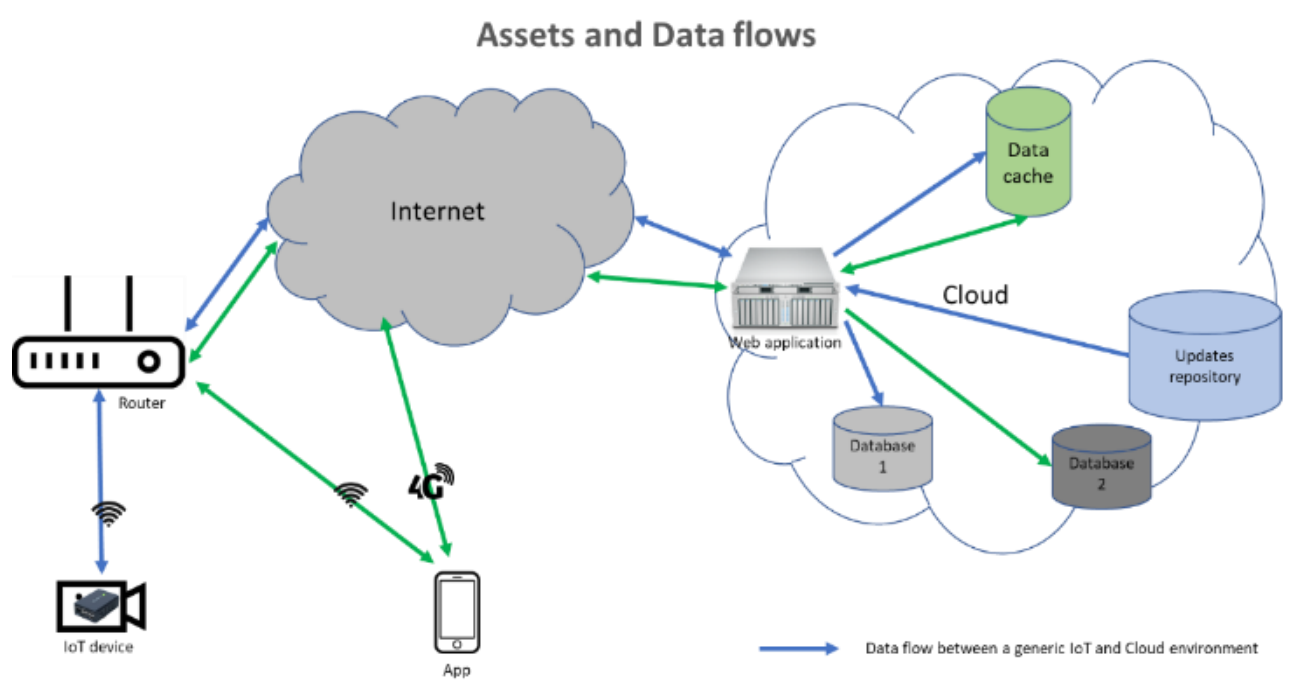

As stressed in the above picture, the generic smart home device is wireless or wired connected to the internet through which it communicates with its cloud databases. The cloud hosts the data the device stores, useful for its aim (the refrigerator data will be the information of the inside products and the temperature, the light lamp data will deal with the actual operation of the devices and the different colouration of the lamp etc..). Moreover, the user interfaces with that device through his smartphone, connected to the internet as well (92% of IoT devices are controlled and regulated by a smartphone or laptop) (Stanislav and Beardsley, 2015).

Analysing the data flow model is crucial to establish the different vulnerabilities and threats which affect the smart home system.

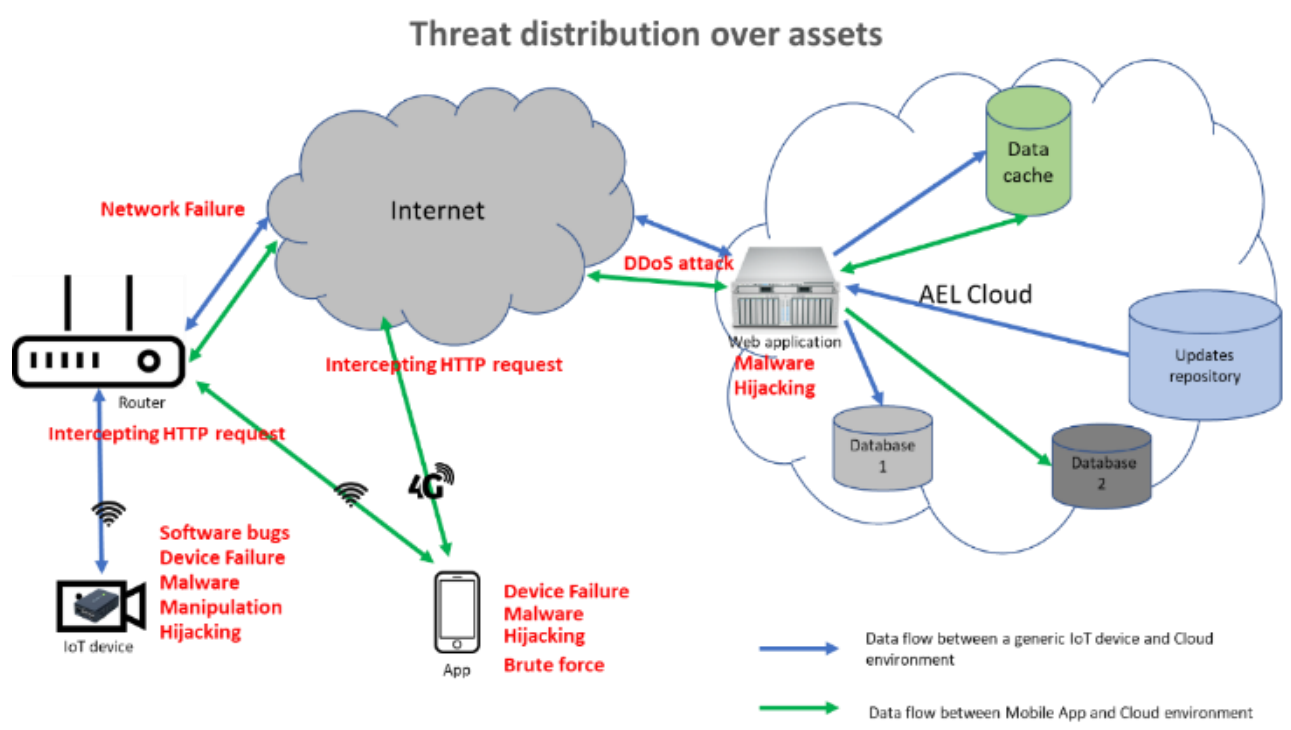

Potentially, an IoT device may suffer, throughout its dataflow sections, the following threats:

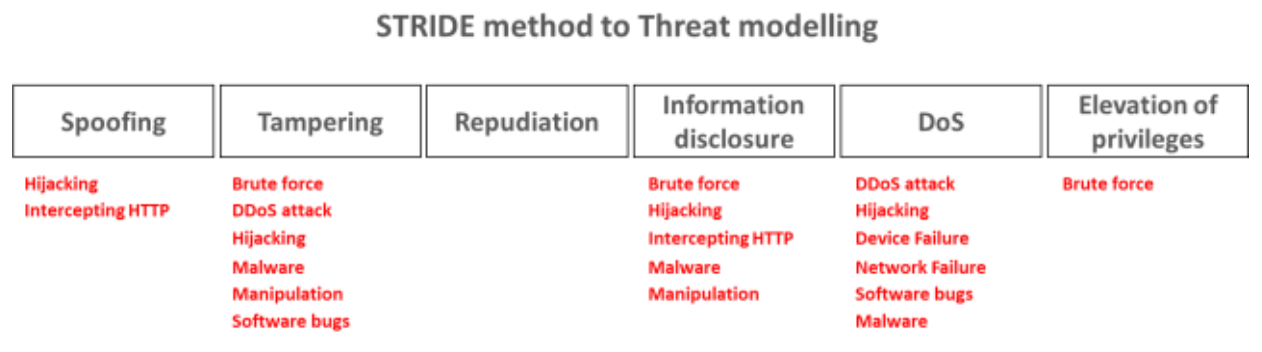

## DEFAULT OR WEAK CREDENTIALS

According to several studies (Enisa.europa.eu, 2020) (Stanislav and Beardsley, 2015), the most critical vulnerability concerning the IoT devices is default or weak credentials. Very often, the user, who lacks security awareness, decides to use default passwords or weak ones for their smart home's solutions just because more likely to be remembered (Enisa.europa.eu, 2020).

Even when there is a certain level of awareness, traditional restrictions (password length and characters impositions), sometimes, generate annoyance and exasperation in users who, then, choose to adopt insecure passwords to overcome these "abrasive charges" (Enisa.europa.eu, 2020).

On the other hand, the malicious actor may be able to either scan the exposed devices utilising online databases of default credentials (Shodan or Insecam) or guess the weak password through dictionary or brute force attack. As a result, the attacker could eventually obtain the credentials, take over the IoT device, and, finally, use it for malicious aims, such as the creation of botnet ("Mirai" constitutes an example). Botnets are usually utilised also for DDoS (Distributed Denial of Service) and cryptographic attacks on different networks (Enisa.europa.eu, 2020).

## CLEARTEXT IN NETWORK COMMUNICATIONS

Other studies underpinned the cleartext in network communications, by which remote communications may not be encrypted. Here, for instance, there may be many passive and active network attacks that can revel sensitive information flowing from and to these IoT devices (Stanislav and Beardsley, 2015).

Almost the majority of all the Smart home devices rely on cloud-based systems which expose APIs from controlling devices over HTTP (Zeng, Mare and Roesner, 2017). As a result, by encompassing an unencrypted connection, all the packets of the internal and external communications, regarding the IoT device, could be potentially sniffed (Stanislav and Beardsley, 2015).

## CHEAP HARDWARE ELEMENTS

Although many IoT devices present cutting-edge features and powerful application, especially for the home automation, these "machines" are usually built with cheap hardware elements. Therefore, cheap firmware and chips usually incorporate built-in vulnerabilities, difficult, even impossible, to detect by operators and owners. In addition, considering those devices which are "always-on and online" (that is to say that keep listening and communicating with internet), IoT is continually and potentially exposed to malware payloads (Team, 2018).

Cheap hardware and weak firmware may be potentially exploitable by the Denial of Service attacks which corrupt the availability of the system by denying users' access. The DoS attack is achieved by flooding the targeted device or network (where the device is hosted) with internet traffic until it crashes, preventing, thus, the access for the authorised users (Us-cert.gov, 2020).

With the widespread of the poorly secured smart home devices, the number of Dos and DDoS attacks has sharply increased and it is estimated that almost the 20% of the IoT attacks finds its roots in these attacks (Crane, 2020) (Cloudflare, 2020).

## III. TECHNIQUES TO ATTACK

This section will present different tools and techniques, used by criminals, to exploit IoT devices.

Here below, the 3 most IoT-related threats the techniques are going to address to:

Brute Force attack on credentials to access the IoT cloud — TYPE: Electronic/Hacking

METHOD: Both default and weak passwords can be easily guessable through a brute force attack.

LIKELIHOOD: Frequent — ASSET INVOLVED: Mobile Application

PROPERTIES:
Confidentiality: High = Rationale: Credentials of the mobile app constitute a high level of privacy
Integrity: Medium = Rationale: Being able to access the app may compromise the integrity of the system

Source references: (Veerendra, 2020) (Metula, 2020)

DDOS — TYPE: Electronic/DoS and DDoS

METHOD: Different systems attack IoT device in order to saturate the cloud service provider. This is done by establishing different connections to the central application, flooding a communication channel or replaying the same communication over and over.

LIKELIHOOD: Occasional — ASSET INVOLVED: Cloud, Web Application

PROPERTIES:
Integrity: High = Rationale: It can jeopardise the functionality of the system
Availability: High = Rationale: it can make the system unavailable
Anonimity: High = Rationale: Attack carried out potentially anonymously

Source reference: (Enisa.europa.eu, 2020)

Intercepting HTTP request from the IoT device to the cloud and from the mobile app to the cloud — TYPE: Electronic/Phishing and Spoofing

METHOD: The attacker sets up a proxy server which intercepts and modifies a HTTP request for 2 different communications: the first one between the IoT device unit and the cloud, via internet, and the second one between the mobile application, which lays on the mobile device, and the cloud, always via internet.

LIKELIHOOD: Probable — ASSET INVOLVED: IoT processor, Router, Mobile device, Web application

PROPERTIES:
Confidentiality: High = Rationale: The attacker wants to corrupt the integrity of the IoT device to steal private data

Source reference: (Stanislav and Beardsley, 2015)

## EXPLOIT IOT THROUGH DEFAULT OR WEAK PASSWORDS

Almost all the IoT devices rely on an application through which they could monitored and/or controlled (the app for a Babycam, amazon Alexa, a smart fridge, washing machine etc..). Similarly, every app, if connected to an online account, asks for a username and password for the authentication. Moreover, usually, a username coincides with the mail.

The attacker, hence, once found the username of the victim, needs his password.

Unfortunately, many IoT products come with default credentials; It is, indeed, estimated that almost 15% of these devices present default password and settings (Ophtek, 2020).

In this scenario, the work for the malicious actor is very simple: googling the name of the device being used by the victim. Online, there are several updated databases of any sort of tech products' default passwords. The most common are: cirt.net, fortypoundhead.com, default-password.info, phenoelit.org, and open-sez.me.

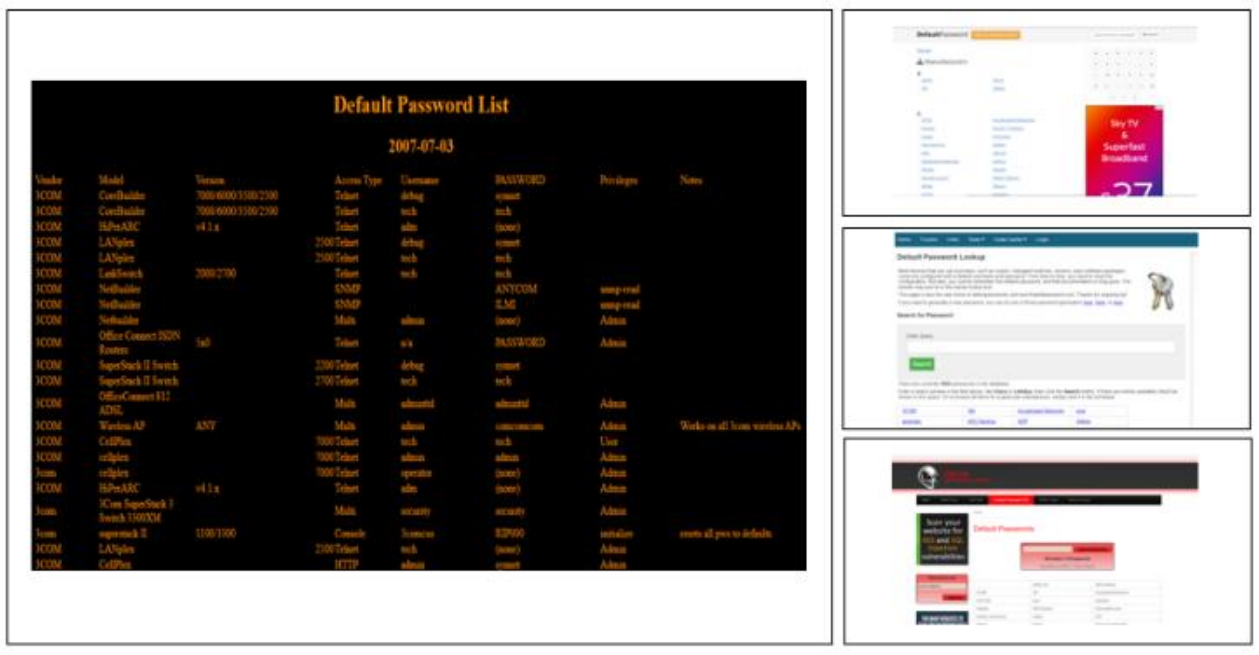

Differently, in case the default password has been changed, the attacker may only try to guess it. A brute force attack is any attempt to discover credentials, hidden web pages or keys used for the communication encryption, by using the "trial and error" approach and "hoping", eventually, to guess correctly. Although it is a dated technique, it is still in use and effective. Length and complexity of passwords play an important role at this step: the longer the password, the less probable guessable (Kaspersky, 2019).

The Daily Swing, a Cyber security newspaper, stated that the first primary issue is the use of weak passwords, which leads to several types of remote compromise (Haworth, 2020). Tracesecurity estimated that 81% of data breaches are facilitated by poor passwords (PCWorld, 2019).

There are different Kali Linux tools for bruteforce attacks. The most known is Hydra, a powerful login cracker that supports many different protocols to attack (kali.org, 2020).

Taking as IoT-device sample an IP smart security camera, the process to follow to run the attack is the following one:

Once got access to the Wlan where the targeted device lays, the attacker is going to scan the network to discover its IP address.

Discovered the IP address of the smart home device, the attacker launches Hydra command that will contain the address of the victim, its username and a wordlist of possible passwords which has been created or downloaded from online resources (Github is the common one also for multilingual wordlists).

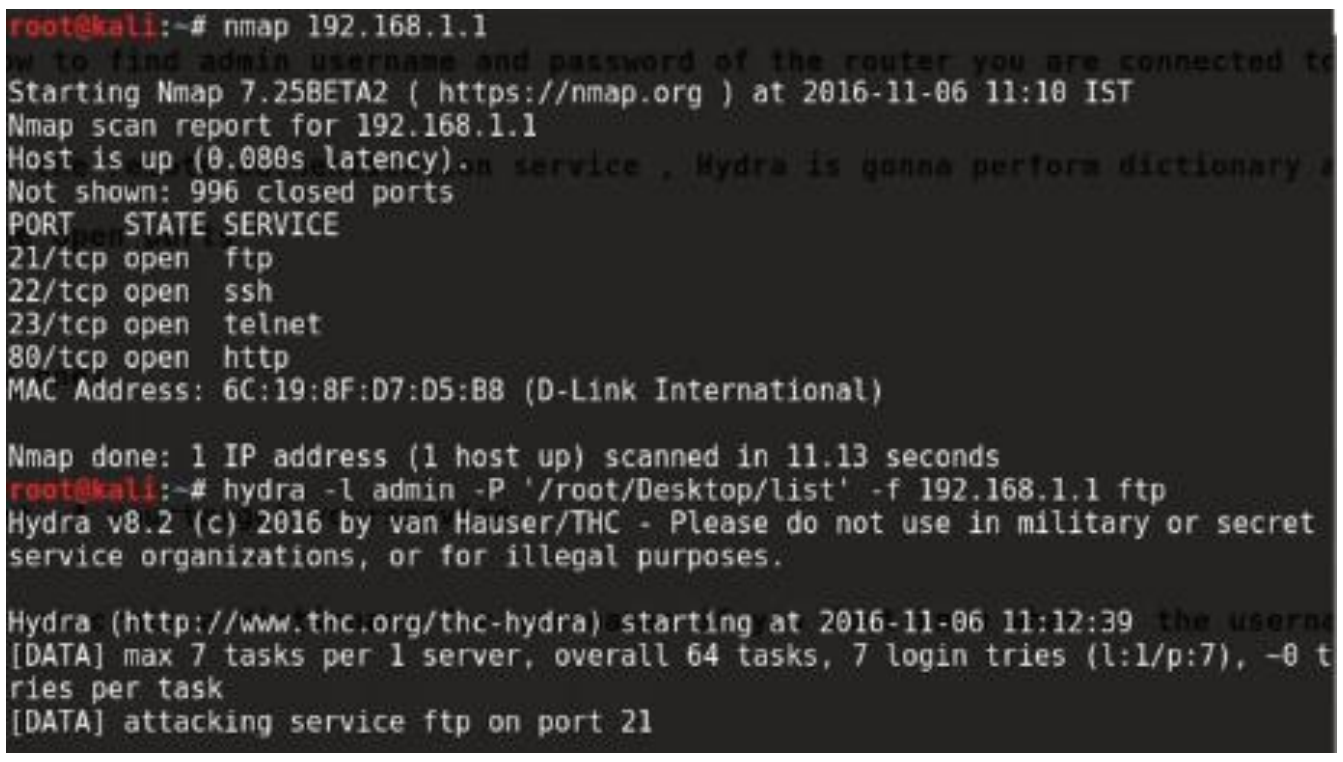

If the attacker is lucky and the password weak, credentials are found.

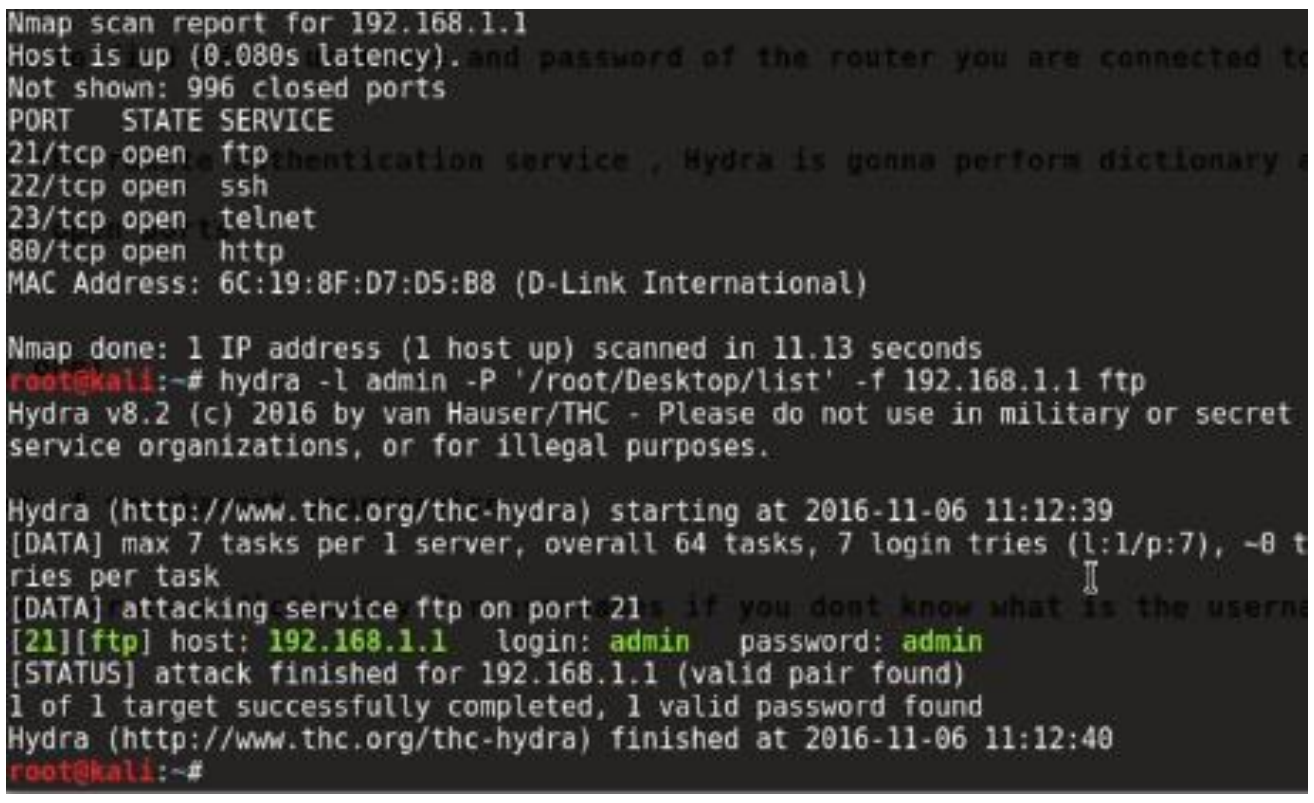

## DOS or DDOS

Regarding the Denial of Service attack, it has already been said that can be facilitated by the cheap material of the hardware and software. However, every system could suffer such attack.

These attacks represent quite the 25% of a country's entire online traffic while they are occurring (Crane, 2019). Moreover, according to Gartner Inc., "IoT devices are notorious for lacking any real IT security or cybersecurity measures, therefore, they're extremely vulnerable to DoS attacks" (Crane, 2019).

A simple but most powerful tool for DOS attacks is Xerxes, committed on Github by Zanyarjamal (Gurubaran, 2018). Unlike Hydra, this tool is not pre-installed on the Kali Linux environment.

Xerxes, as other tools, keeps putting heavy loads on the HTTP server in order to exhaust its resources hence make the device crash.

This technique can be applied to every system (web site, device, network, application), whatever nature it belongs to. Both online servers and devices are usually targeted by malicious actors to deny the availability to the victims. In the following demonstration, an online hacking game website will be used not to cause any physical damage.

Once installed the tool, the attacker starts the attack:

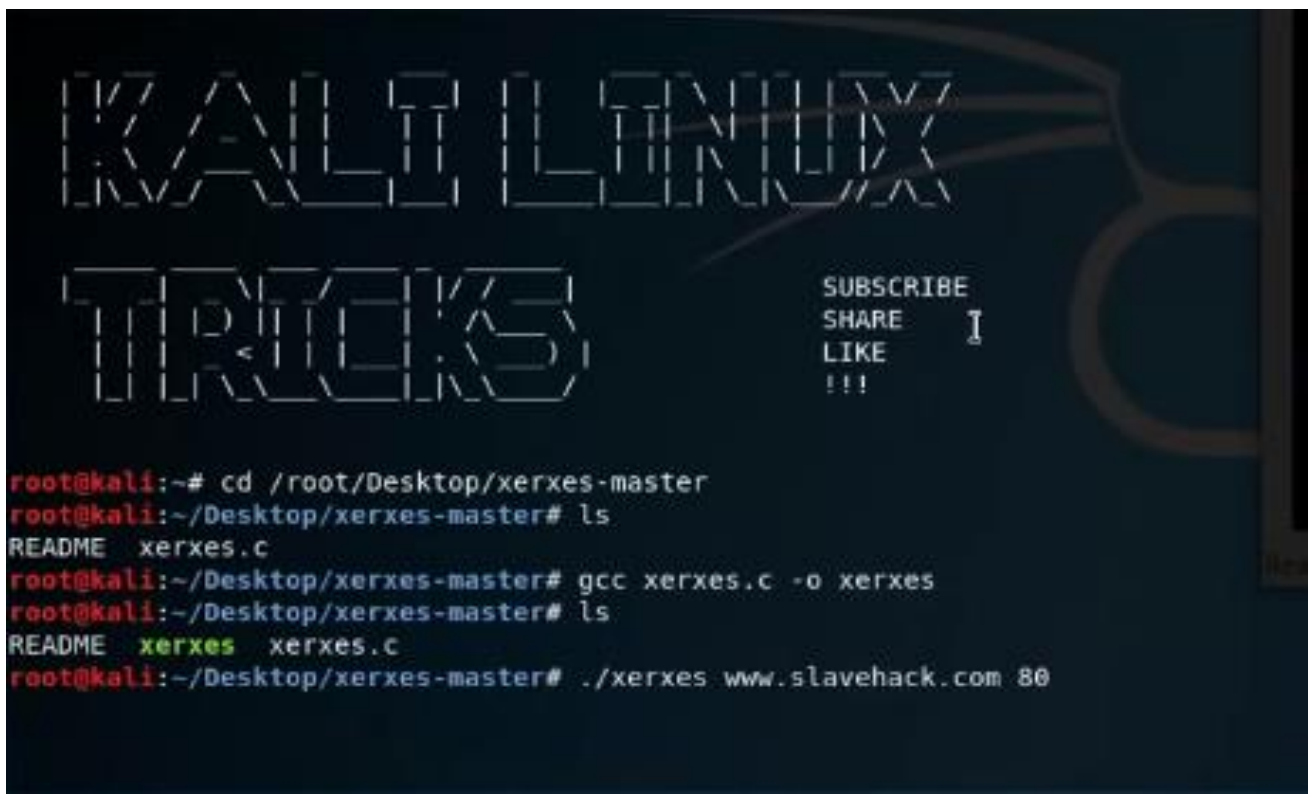

Once the command is executed, the system starts sending request through the port 80 (HTTP) to the IP or URL address (in this case the URL of the website Slavehack.com).

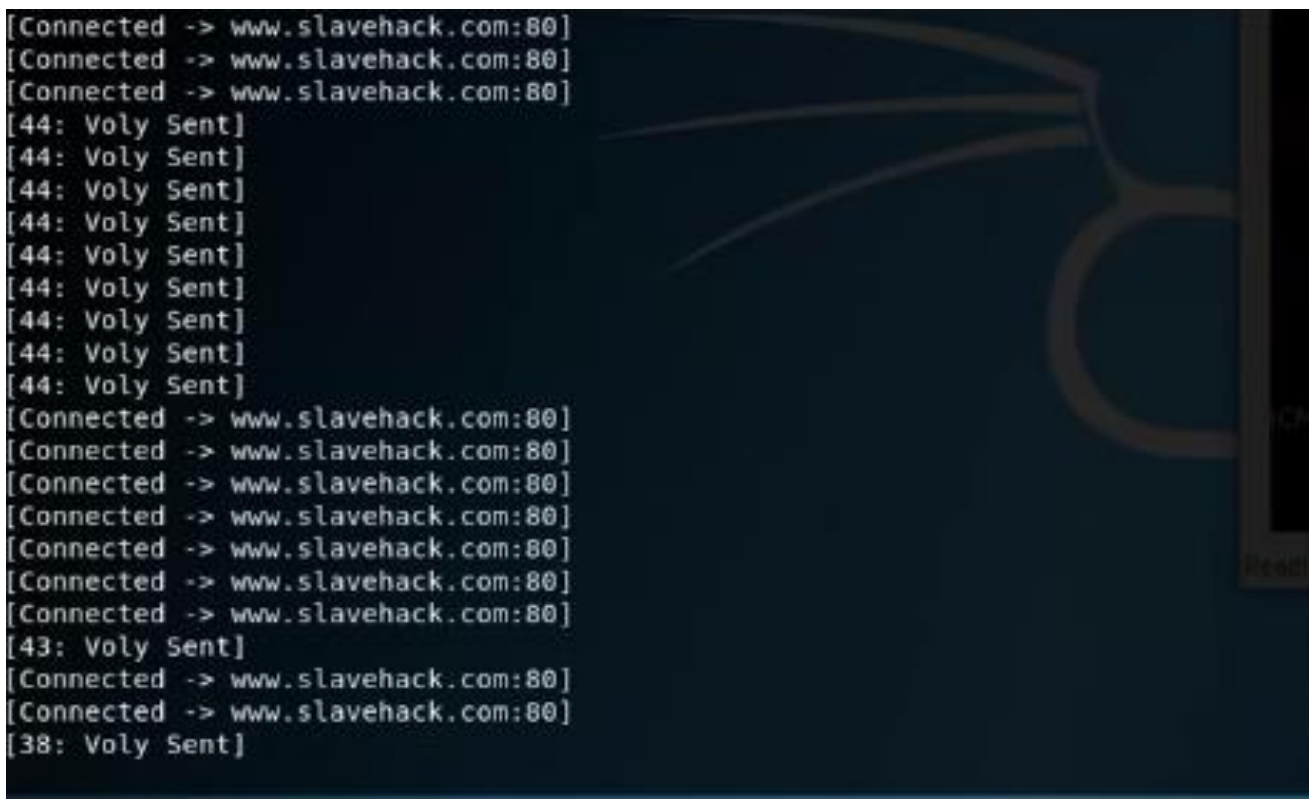

Once saturated, the system crashes preventing, hence, the victim from using it.

## INTERCEPTING THE HTTP REQUESTS FROM AND TO THE IOT DEVICES

Many smart home systems interface with the "external world" through a non-secure protocol, such as HTTP. This could lead to eavesdropping of the traffic between them and users (Enisa.europa.eu, 2020).

The attacker sets a proxy server aimed to intercept and modify the HTTP requests. Proxy servers may be either a computer or an app hosted in the computer; it could sometimes also be somewhere between the online server and the client (Kayode and Tosun, 2018).

The proxy server behaves as intermediary and helps to forward the requests of the host towards the desired target host. If used during HTTP communication, the proxy would have access to the transmitted unencrypted data. As a result, those devices which transmit data in cleartext will easily expose sensitive user's information. The malicious actor may thus get access to user's data through this method (Kayode and Tosun, 2018).

Wireshark is a popular open source graphical user interface (GUI) tool for analysing packets within a network. This tool is particularly useful for (Nasi, 2020):

- Capture live packet data from a network interface.
- Open files containing packet data captured with tcpdump/WinDump, Wireshark, and a number of other packet capture programs.
- Import packets from text files containing hex dumps of packet data.

Once opened the tool, which is pre-installed in all Debian-base OS, the scan of the network is initiated.

All the unencrypted communications which pass in the network are eavesdropped by Wireshark and all plain text resources are given, here below it is shown how the plain text credentials of this IoT device are sniffed by this tool:

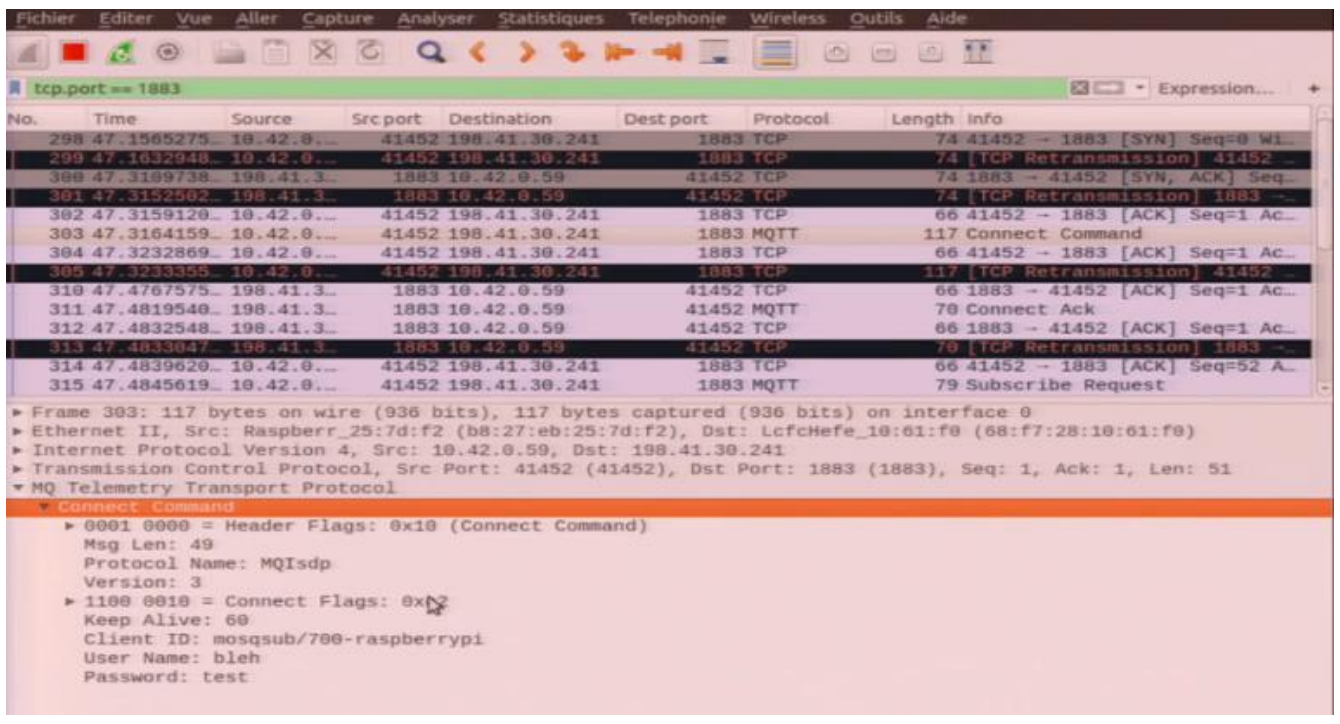

## IV. COUNTERMEASURES

The attacks explained in the previous section can be mitigated by countermeasures aimed to protect the users' data and the integrity of the IoT devices.

The principal vulnerability is the default and weak passwords; when it comes to choosing a strong password, primarily and whenever possible, it is highly recommended to use strong authentication mechanisms (for instance, based on challenge-response authentication, with an SSH signature) (Enisa.europa.eu, 2020). Furthermore, it is also essential to make sure that the authentication mechanism prevents users from creating weak credentials (such as keywalk passwords, that is, passwords based on adjacent keyboard keys, e.g. 'qwertyuiop', or obvious ones, like 'Aa12345!'). There are plenty of international authorities (NIST, 2020) that have created guidelines for this purpose, but essentially, users should be given freedom to be creative and create custom passwords that are both secure and user-friendly, without too many format restrictions (Enisa.europa.eu, 2020).

Online databases of exposed or default credentials (https://haveibeenpwned.com/) are a good resource to avoid the use of weak authentication mechanisms. This would hinder the malicious actions of the attacker, since the tools available would be less effective to guess weak or default passwords (Enisa.europa.eu, 2020). Another recommendation is to implement security mechanisms like multiple-factor authentication for application access or mechanisms forcing to change or set up a new password before using the device for the first time (Enisa.europa.eu, 2020).

Regarding DoS and DDoS attacks, unfortunately, "there is no silver bullet" (Weagle, 2020), but there are some precautions which should be taken to protect against them (Weagle, 2020).

The selection of smart home devices from reputable companies which are committed to selling secure products is the primary tip. In addition, users should also regularly check for firmware updates to make sure the system is not compromised. (Weagle, 2020) (Enisa.europa.eu, 2020).

A possible technical solution for these attacks is the SDN (Software-defined networking), a network security management used throughout different areas, such as, e-health, business and smart homes as well (Tabassum and Lebda, 2019).

Giotis et al. (2014) tested the capability of an SDN architecture to protect the IoT devices against DOS attacks. They found out that SDN was very efficient to detect such attacks and did not cause an overhead to the controller (Giotis et al., 2014).

Another solution was suggested to be the IDS but this is going to be treated at the end of this section (Papamartzivanos, Gomez Marmol and Kambourakis, 2019).

Finally, the mitigation of eavesdropping and sniffing.

While data is transmitted from the device to the server, it can be hijacked. An attacker may sniff the communication traffic and alter it leading to serious privacy concerns. However, there are some possible solutions to mitigate the risk and protect the communication as much as possible. (Yousuf, Mahmoud, Aloul and Zualkernan, 2015) (Tabassum and Lebda, 2019).

To mitigate both DDoS/DoS and eavesdropping, it could be used the IDS (Intrusion detection system) which is a program that manages to discover doubtful activities in a network environment and identify unauthorised access to the IoT device (Tabassum and Lebda, 2019). The IDS, although it is sometimes unable to detect new attacks and slow in its response, may eventually constitute a countermeasure against Cyberattacks (Papamartzivanos, Gomez Marmol and Kambourakis, 2019) (Tabassum and Lebda, 2019).

As shown in the "intercepting the http request" example, whenever a communication uses an HTTP protocol, there is a possible leakage of the users' privacy. Credentials in plain text are sniffed and the confidentiality exploited (Tabassum and Lebda, 2019).

The primary solution is to set on those devices a security communication encompassing the HTTPS protocol (Tabassum and Lebda, 2019). Having an encrypted communication may defend users' confidentiality.

In addition, to foster the security of IoT devices, it could be implementing a VPN (Rottigni, 2019).

The function of VPNs is not just encrypting data, they also cover users' geographical location and IP address. Protecting this data helps prevent third parties, whether hackers, Internet Service Providers, government agencies, or others who might try to gather information about your activities. This means that other entities cannot infiltrate an IoT device and start eavesdropping or leaking confidential information (Rottigni, 2019).

VPN can, hence, protect users against different and common IoT-related cyberattacks: eavesdropping and botnets (Rottigni, 2019).

## V. CONCLUSION AND FUTURE RECCOMENDATIONS

The IoT (Internet of Things) has become widely popular in the domestic environments. People are renewing their homes into smart homes; however, the privacy concerns of owning many Internet connected devices with always-on environmental sensors remain insufficiently addressed. Throughout this paper, different threats and vulnerabilities of smart home devices have been found and described. Default and weak passwords are the primary concern while cheap materials and hardware and unencrypted communication could lead to a leakage of the integrity and the confidentiality of such devices.

Solutions and countermeasures have been also provided: choosing a strong password, strong authentication mechanisms, check online databases of exposed or default credentials to mitigate the first threat; a selection of smart home devices from reputable companies and the implementation of the SDN for the DoS threat; and finally IDS, HTTPS protocol and VPN for eavesdropping.

However, as stressed in the introduction, it is expected that the number of IoT devices will increase. A lack of technical support constitutes another big challenge for the smart home (Lin and Bergmann, 2016). Users “feel bothered by tedious, repetitive and error-prone manual tasks” for setting these devices and eventually fixing them. This could pose, indeed, a further security risk. Therefore, for a more-secure and successful implementation of IoT devices, an auto-configuration approach should be analysed since it could both ease their installation/maintenance and enhance the security in the self configuration step (Lin and Bergmann, 2016).

The system created by Lin et al. (2016) implies that, whenever a new device is connected to the network, the gateway interrogates a trusted and secure web server, according to the device ID, with the aim to find the details of such device, e.g. its functionality, what encryption and networking protocols it understands or eventual firmware updates. Differently from others, this approach ensures, with just a web server and a simple device ID, availability, integrity, and confidentiality of the personal data (Lin and Bergmann, 2016).

The topic of the next project, indeed, will be the secure auto-configuration approach which could lead to a “next generation” of IoT security.

## VI. REFERENCES

Zeng, E., Mare, S. and Roesner, F. (2017). End User Security and Privacy Concerns with Smart Homes. Thirteenth Symposium on Usable Privacy and Security, [online] pp.65-80. Available at: https://www.usenix.org/conference/soups2017/technical-sessions/presentation/zeng [Accessed 4 Mar. 2020].

Ncsc.gov.uk. (2019). Smart devices: using them safely in your home. [online] Available at: https://www.ncsc.gov.uk/guidance/smart-devices-in-the-home [Accessed 4 Mar. 2020].

Apthorpe, N., Reisman, D. and Feamster, N. (2019). A Smart Home is No Castle: Privacy Vulnerabilities of Encrypted IoT Traffic. Proceedings on Privacy Enhancing Technologies, 2019(3), pp.128-148.

Apthorpe, N., Huang, D., Reisman, D., Narayanan, A. and Feamster, N. (2019). Keeping the Smart Home Private with Smart(er) IoT Traffic Shaping. Proceedings on Privacy Enhancing Technologies, 2019(3), pp.128-148.

Enisa.europa.eu. (2020). Good Practices for Security of Internet of Things in the context of Smart Manufacturing. [online] Available at: https://www.enisa.europa.eu/publications/good-practices-for-security-of-iot [Accessed 4 Mar. 2020].

Angrishi, K. (2017). Turning Internet of Things(IoT) into Internet of Vulnerabilities (IoV) : IoT Botnets. Cornell University. [online] Available at: https://arxiv.org/abs/1702.03681v1 [Accessed 5 Mar. 2020].

Stanislav, M. and Beardsley, T. (2015). HACKING IoT: A Case Study on Baby Monitor Exposures and Vulnerabilities. [online] Available at: https://www.rapid7.com/globalassets/external/docs/Hacking-IoT-A-Case-Study-on-Baby-Monitor-Exposures-and-Vulnerabilities.pdf [Accessed 11 Nov. 2019].

Team, F. (2018). IoT DoS Attacks | Hacked IoT Devices Can Lead To Massive DoS Attacks. [online] Finjan Blog. Available at: https://blog.finjan.com/iot-dos-attacks/ [Accessed 5 Mar. 2020].

Us-cert.gov. (2020). Understanding Denial-of-Service Attacks | CISA. [online] Available at: https://www.us-cert.gov/ncas/tips/ST04-015 [Accessed 5 Mar. 2020].

Crane, C. (2020). 20 Surprising IoT Statistics You Don’t Already Know - Security Boulevard. [online] Security Boulevard. Available at: https://securityboulevard.com/2019/09/20-surprising-iot-statistics-you-dont-already-know/ [Accessed 5 Mar. 2020].
Cloudflare. 2020. How To Ddos | Dos And Ddos Attack Tools. [online] Available at: <https://www.cloudflare.com/> [Accessed 29 April 2020].

Ophtek, L., 2020. Default Passwords: The Biggest Weakness In Iot Security - Ophtek. [online] Ophtek. Available at: <https://www.ophtek.com/default-passwords-biggest-weakness-iot-security/> [Accessed 29 April 2020].

Kaspersky. 2019. What's A Brute Force Attack?. [online] Available at: <https://www.kaspersky.com/resource-center/definitions/brute-force-attack> [Accessed 29 April 2020].

Haworth, J., 2020. OWASP: Weak Passwords Are Biggest Threat To Iot Security. [online] The Daily Swig | Cybersecurity news and views. Available at: <https://portswigger.net/daily-swig/owasp-weak-passwords-are-biggest-threat-to-iot-security> [Accessed 29 April 2020].

PCWorld. 2019. Hacked Passwords Cause 81% Of Data Breaches - Media Releases - PC World Australia. [online] Available at: <https://www.pcworld.idg.com.au/mediareleases/29642/hacked-passwords-cause-81-of-data-breaches/> [Accessed 29 April 2020].

kali.org. 2020. Hydra Package Description. [online] Available at: <https://tools.kali.org/password-attacks/hydra> [Accessed 29 April 2020].

Crane, C., 2019. The 15 Top Ddos Statistics You Should Know In 2020. [online] Cybercrime Magazine. Available at: <https://cybersecurityventures.com/the-15-top-ddos-statistics-you-should-know-in-2020/> [Accessed 29 April 2020].

Gurubaran, S., 2018. Kali Linux Tutorial - Most Powerful Dos Tool XERXES. [online] GBHackers On Security. Available at: <https://gbhackers.com/xerxes-kali-linux-tutorial/> [Accessed 29 April 2020].

Kayode, O. and Tosun, S., 2018. Analysis of IoT Traffic using HTTP Proxy.

NIST, 2020. Digital Identity Guidelines. [online] Pages.nist.gov. Available at: <https://pages.nist.gov/800-63-3/sp800-63b.html> [Accessed 29 April 2020].

Weagle, S., 2020. The Iot Makes It Easier To Launch Massive Ddos Attacks - Corero. [online] Corero. Available at: <https://www.corero.com/blog/the-iot-makes-it-easier-to-launch-massive-ddos-attacks/> [Accessed 29 April 2020].

Tabassum, A. and Lebda, W., 2019. SECURITY FRAMEWORK FOR IOT DEVICES AGAINST CYBER-ATTACKS. Department of Computer Science and Engineering, Qatar University,.

Giotis, K., Argyropoulos, C., Androulidakis, G., Kalogeras, D. and Maglaris, V., 2014. Combining OpenFlow and sFlow for an effective and scalable anomaly detection and mitigation mechanism on SDN environments. Computer Networks, 62, pp.122-136.

Papamartzivanos, D., Gomez Marmol, F. and Kambourakis, G., 2019. Introducing Deep Learning Self-Adaptive Misuse Network Intrusion Detection Systems. IEEE Access, 7, pp.13546-13560.

Yousuf, T., Mahmoud, R., Aloul, F. and Zualkernan, I., 2015. Internet of Things (IoT) Security: Current Status, Challenges and Countermeasures. International Journal for Information Security Research, 5(4), pp.608-616.
Nasi, M., 2020. Wireshark, Una Breve Guida All'uso. [online] IlSoftware.it. Available at: <https://www.ilsoftware.it/articoli.asp?tag=Wireshark-una-breve-guida-all-uso_18866> [Accessed 29 April 2020].

Rottigni, R., 2019. How A VPN Can Enhance Your IOT Devices Security - Readwrite. [online] ReadWrite. Available at: <https://readwrite.com/2019/02/06/how-a-vpn-can-enhance-your-iot-devices-security/> [Accessed 29 April 2020].
Lin, H. and Bergmann, N., 2016. IoT Privacy and Security Challenges for Smart Home Environments. Information, 7(3), p.44.